\documentclass[%
 reprint,
 amsmath,amssymb,
 aps,
]{revtex4-2}

\usepackage{dcolumn}
\usepackage{bm}

\usepackage[utf8]{inputenc}
\usepackage[T1]{fontenc}
\usepackage{multirow}
\usepackage{subcaption}
\usepackage{tabularx}
\usepackage{graphicx,caption}
\usepackage{amsfonts,amsmath,amssymb}
\usepackage{comment}
\usepackage{lineno}
\usepackage{hyperref}
\usepackage{cleveref} 

\newcommand{\be}{\begin{equation}}
\newcommand{\ee}{\end{equation}} 
\newcommand{\bea}{\begin{eqnarray}}
\newcommand{\eea}{\end{eqnarray}}

\renewcommand{\bf}[1]{\textbf{#1}} 

\newcommand{\ra}{\rightarrow}
\newcommand{\f}[2]{\frac{#1}{#2}}

\newcommand{\ccup}[1]{\left\{#1\right\}}

\newcommand{\bup}[1]{\left(#1\right)}
\newcommand{\rup}[1]{\left[#1\right]}

\newcommand{\Exp}{\mathbb{E}}

\renewcommand{\ref}[1]{[\ref{#1}]}
  
\usepackage{algorithm,algorithmic} 
\usepackage[algoruled,algo2e,ruled]{algorithm2e}  
\usepackage{sidecap}
\usepackage{boldline}
\usepackage{setspace}
\usepackage{xcolor}

\usepackage[table]{}
\usepackage{array}
\usepackage{booktabs}
\usepackage{adjustbox}  
 \usepackage{natbib}
\usepackage{float}

\usepackage{tabu}
\usepackage{tikz}

\usepackage[normalem]{ulem} 
 
\newcommand{\jcrep}{\mbox{{\small JointCRep}}}  
\newcommand{\ACD}{\mbox{{\small ACD}}}
\newcommand{\ajcrep}{\mbox{{\small CRAD}}} 
\newcommand{\vimure}{\mbox{{\small VIMuRe}}}

\usepackage{graphicx,caption} 
\usepackage{cleveref}
\crefname{equation}{Eq.}{Eqs.}
\crefname{section}{Sec.}{Secs.}
\crefname{figure}{Fig.}{Figs.}

\begin{document}

\preprint{APS/123-QED}

\title{Anomaly, reciprocity, and  community detection in networks}

\author{Hadiseh Safdari}
\email{hadiseh.safdari@tuebingen.mpg.de} 
\author{Martina Contisciani}%
\email{martina.contisciani@tuebingen.mpg.de}
\author{Caterina De Bacco}%
\email{caterina.debacco@tuebingen.mpg.de} 
\affiliation{Max Planck Institute for Intelligent Systems, Cyber Valley, Tuebingen 72076, Germany}

\date{\today}

\begin{abstract}
Anomaly detection algorithms are a valuable tool in network science for identifying unusual patterns in a network. These algorithms have numerous practical applications, including detecting fraud, identifying network security threats, and uncovering significant interactions within a dataset. In this project, we propose a probabilistic generative approach that incorporates community membership and reciprocity as key factors driving regular behavior in a network, which can be used to identify potential anomalies that deviate from expected patterns. We model pairs of edges in a network with exact two-edge joint distributions. As a result, our approach captures the exact relationship between pairs of edges and provides a more comprehensive view of social networks. Additionally, our study highlights the role of reciprocity in network analysis and can inform the design of future models and algorithms. We also develop an efficient algorithmic implementation that takes advantage of the sparsity of the network. 

\end{abstract}

\maketitle


\section{\label{sec:Intro}Introduction}
Anomaly detection algorithms are a crucial tool in the study of networks. These algorithms are designed to identify unusual or unexpected patterns in the data, which can provide valuable insights into the structure and function of a network \cite{Akoglusurvey2015, Hodge:2004aa}. For instance, anomalous edges in a network may indicate the presence of a structural flaw or a potential problem, such as a vulnerability to attack. By detecting and analyzing these anomalies, we can gain a better understanding of the network and potentially identify ways to improve its performance or security \cite{Bhuyan2014}. In addition, anomaly detection algorithms can be used to monitor networks in real-time, allowing researchers to quickly identify and respond to potential issues as they arise. 

Anomalies are often difficult to define precisely because they can vary depending on the context and the system being analyzed \cite{Hawkins1980}. For example, in a network of online transactions, an anomaly could be a sudden spike in the number of transactions coming from a single user \cite{Aleskerov1997}. In this case, the regular behavior in the system would be the typical number of transactions coming from a single user, and any deviation from this pattern would be considered as anomaly.
Hence, one of the main obstacles in detecting anomalies in networks is determining what is considered ``normal" (or ``regular'') behavior. To overcome this challenge, we must create a null model which is a realistic representation of the network data. This null model provides a standard against which we can compare the network data and identify anomalies. \\
Relevant approaches to address this problem include statistics-based methods, which fit a  statistical model to the network data \cite{Chandola2009, Eskin2000}. Among these, generative models \cite{goldenberg2010survey,bojchevski2018bayesian, SAVAGE2014} make assumptions about the processes that drive network formation and evolution to generate synthetic network data. By using these approaches, we can define null models that are tailored to the specific characteristics of the network under study. This is the approach we take here.

In this work, we focus on plain networks, which only contain information about the presence or absence of connections between individuals, and do not include any additional information.  One approach to perform anomaly detection in these binary and single-layer networks is to use the structure of the graph to identify patterns and detect deviations from them \cite{Akoglusurvey2015}. These structural patterns can be  divided into two categories: patterns based on the \textit{overall structure} of the graph, and patterns based on the \textit{community structure} of the graph. Methods in the first category rely on the global properties of the graph \cite{Henderson2010}, such as the distribution of node degrees or the overall connectivity of the network. On the other hand, methods in the second category  perform anomaly detection by focusing on the local properties of the graph, such as the membership of nodes in communities \cite{nikulin2012a,Perozzi2018}. Hence, with the second approach, we assume that the null model reflects a community structure that can be identified through latent variables, a process known as community detection task \cite{fortunato2016community}. Thus, by considering the community structure, anomalous behavior can be determined in this context.  For example, a friendship between two individuals from different groups, such as high school classmates and college classmates, could be considered anomalous.    We recently developed a model (\ACD) that performs anomaly detection by using community structure \cite{SafdariJBD2022}, where anomalous edges are those that deviate from regular patterns determined by community structure. As a result, this model outputs both node memberships and edge labels identifying them as legitimate or anomalous.  

Accurately identifying anomalies is deeply connected with the chosen null model determining what regular patterns are. As a consequence, it is important to consider other possible mechanisms for tie formation, beyond community structure. For instance, reciprocity, another fundamental structural feature in networks \cite{wasserman1994social,deBacco2021vimure,Molm2010}, refers to the mutual exchange of resources or actions between individuals or groups. This can include actions such as returning a favor, sharing information or resources, or collaborating on a project. For example, in a social network, if two individuals consistently like and comment on each other's posts, this could be considered reciprocity. In a business network, if two companies frequently refer customers to each other, this could also be considered reciprocity. Mathematically, it is calculated as the ratio of the number of reciprocated edges to the total number of edges in the graph.
Recent works \cite{contisciani2021JointCrep,safdari2020generative}  have shown that including reciprocity effects in the modeling of community patterns results in more accurate and expressive generative models.  This has the potential to improve the performance of an anomaly detection model for networks  as well.
 
In this work, we develop a probabilistic generative model that we refer to as Community Reciprocity Anomaly Detection (\ajcrep) algorithm,  that performs anomaly detection by proposing a null model based on both community structure and reciprocity. Intuitively, our model regards as regular ties  those who follow the group membership and reciprocity effects, and as anomalous ties those whose formation process is not aligned with these two mechanisms. Notice that node memberships, reciprocity effect, and anomalous edges are all unknown processes. Our model is able to infer them from data by representing them as latent variables in a probabilistic model.

More specifically,  we model the existence of  ties between pairs of nodes using a bivariate Bernoulli distribution. This has the crucial statistical property that independence and uncorrelatedness of the component random variables are equivalent \cite{dai2013multivariate}, which facilitates the derivation of a closed-form joint distribution of a pair of edges.  Furthermore, both the marginal and conditional distributions  are Bernoulli distributions, enabling closed-form analytical expressions. This facilitates downstream analysis and also improves model performance, as shown in \cite{contisciani2021JointCrep}.

\section{\label{sec:model}The Model}
We are given an adjacency matrix, $\textbf{A}$ as our observed data, with entries indicating the presence or absence of an edge from node $i$ to node $j$, represented by $A_{ij}=1$ or $A_{ij}=0$, respectively.  Pairs of directed edges between two nodes $(i,j)$ are defined as $A_{(ij)}=(A_{ij}, A_{ji})$. We consider binary data, thus $A_{(ij)} \in \ccup{0, 1}^2 = \ccup{0, 1} \times  \ccup{0, 1}$,  and directed networks, i.e., in general $A_{ij}\neq A_{ji}$. We aim at classifying any such pair as either regular or anomalous, accounting for community structure and reciprocity effects.  For this, we introduce a Bernoulli random variable that represents the binary label of  being anomalous or not as a random variable:
\begin{equation}
\sigma_{(ij)} \sim \text{Bern}(\mu)  \ , \label{eqn:sigma_prior}    
\end{equation}
where $\sigma_{(ij)} =0,1$ if the pair $A_{(ij)}$ is regular or anomalous, respectively.    In this work we assume that edges between any pair of nodes must be either anomalous or regular. Mathematically, this means that the matrix $\boldsymbol \sigma$ with entries $\sigma_{ij}$ is symmetric, i.e., $\sigma_{ij} =\sigma_{ji}$. These latent variables must be learned from data, as anomalies are not known in advance. They also determine the mechanism from which the pair of edges are drawn. The hyper-parameter $\mu \in \rup{0,1}$ controls the prior distribution of $\sigma_{(ij)}$.
 
With these main ingredients in mind, we can proceed to characterize the joint probability distribution of pairs of edges. Assuming to know the label $\sigma_{(ij)}$ for a given pair of edges, we denote the pair joint probability $p_{nm}^{(\ell)}=P^{(\ell)}(A_{ij}=n, A_{ji}=m)$, where $n,m\in\{0,1\}$ and $\ell \in \ccup{r,a}$ denotes the label being regular or anomalous, respectively. 
 We then consider the joint probability distribution of a pair of edges as a bivariate Bernoulli distribution:
\begin{small}
\begin{align}\label{eqn:P_joint}
&P(A_{(ij)}, \sigma_{(ij)}) =P(A_{ij},A_{ji},\sigma_{(ij)}) = P(A_{ij},A_{ji}|\sigma_{(ij)})\, P(\sigma_{(ij)}) \nonumber \\  \nonumber \\
&= P^{(a)}(A_{ij},A_{ji}|\theta_{a})^{\sigma_{(ij)}} \,P^{(r)}(A_{ij}, A_{ji}| \theta_{r})^{1-\sigma_{(ij)}}\, P(\sigma_{(ij)}|\mu)  \nonumber\\  \nonumber \\
&=  \rup{[p^{(a)}_{11}]^{A_{ij}A_{ji}}[p^{(a)}_{10}]^{A_{ij}(1-A_{ji})}[p^{(a)}_{01}]^{(1-A_{ij})A_{ji}}[p^{(a)}_{00}]^{(1-A_{ij})(1-A_{ji})}}^{\sigma_{(ij)}}   \nonumber \\ \nonumber \\
&\times \rup{[p^{(r)}_{11}]^{A_{ij}A_{ji}}[p^{(r)}_{10}]^{A_{ij}(1-A_{ji})}[p^{(r)}_{01}]^{(1-A_{ij})A_{ji}}[p^{(r)}_{00}]^{(1-A_{ij})(1-A_{ji})}}^{1-\sigma_{(ij)}}\,\nonumber \\ \nonumber \\ & \times  \mu^{\sigma_{(ij)}} \, (1-\mu)^{1-\sigma_{(ij)}}\ ,
\end{align}
\end{small}
where $\theta_r$ and $\theta_a$   denote parameters specific to the two distributions $P^{(r)}$ and $P^{(a)}$. 
The parameters $p_{nm}^{(\ell)}$ must satisfy $\sum_{n,m=0,1}p_{nm}^{(\ell)}=1$ to have valid probability density functions.

Following the notation as in  \cite{contisciani2021JointCrep,dai2013multivariate}, we can rewrite  the full joint probability density function in \cref{eqn:P_joint}, as the product, 
\begin{widetext}
\begin{align}\label{eqn:PT_joint}
P(\boldsymbol A,\boldsymbol \sigma) &= \prod_{(i,j)}\;   \Bigg[ \f{ \exp \ccup{  A_{ij} f^{(a)}_{ij}+A_{ji} \,f^{(a)}_{ji} + A_{ij}A_{ji}\,J_{(ij)} ^{(a)}}}{Z^{(a)}_{(ij)}}\times  \mu  \Bigg]^{\sigma_{(ij)}} \Bigg[ \f{ \exp \ccup{ A_{ij}f_{ij}^{(r)}  + A_{ji}f_{ji}^{(r)}+ A_{ij}A_{ji}\,J_{(ij)}^{(r)}}}{Z_{(ij)}^{(r)}}\times  (1-\mu)  \Bigg]^{1-\sigma_{(ij)}} \ ,
\end{align}
\end{widetext} 
where $p^{(\ell)}_{00}=1/Z_{(ij)}^{(\ell)}$, and $Z_{(ij)}^{(\ell)}$  is the normalization constant  for the regular or anomalous edges, for $\ell \in \ccup{r,a}$; $f_{ij}^{(\ell)}, f_{ji}^{(\ell)}$, and $J_{(ij)}^{(\ell)}$ are the natural parameters of their density functions. The interaction term $J_{(ij)}^{(\ell)}$ appears  in order to capture reciprocity. It allows to have a joint pair distribution $P(A_{ij},A_{ji}|\sigma_{(ij)})$ that is not simply the product of two independent distributions $P(A_{ij}|\sigma_{(ij)})\,\times P(A_{ji}|\sigma_{(ij)})$, as it is usually assumed in cases where reciprocity (or other properties involving more than on variable) is not taken into account explicitly.  \\
These parameters can be expressed in terms of the probability of occurrence of edges as follows:
\bea\label{eqn:ftot} 
& f_{ij}^{(\ell)}  = \log\bup{\f{p^{(\ell)}_{10}}{p^{(\ell)}_{00}}}  \ , \
f_{ji}^{(\ell)} = \log\bup{\f{p^{(\ell)}_{01}}{p^{(\ell)}_{00}}} \ ,  \nonumber \\ \nonumber \\
&  J_{(ij)}^{(\ell)} = \log\bup{\f{p^{(\ell)}_{11}p^{(\ell)}_{00}}{p^{(\ell)}_{10}p^{(\ell)}_{01}}}  \ ,\ \ell = \ccup{r,a}\,.   \\ \nonumber
\eea

%
%
%

\definecolor{deg2}{rgb}{0.993482, 0.765499, 0.156891}
\definecolor{deg3}{rgb}{0.898984, 0.420392, 0.363047}
\definecolor{deg4}{rgb}{0.660374, 0.134144, 0.588971}
\definecolor{deg5}{rgb}{0.293478, 0.010213, 0.62949}

\definecolor{ligre}{rgb}{.522, .796, .980}

\definecolor{darkgreen}{rgb}{0.0, 0.5, 0.0}
\definecolor{darkblue}{rgb}{0, 0.24, 0.64}
\definecolor{darkorange}{rgb}{1, 0.45, 0.0}
\definecolor{yellowgreen}{rgb}{0.6, 0.8, 0.4}

\begin{figure}[b]
  \captionsetup[subfigure]{justification=centering}
  \begin{subfigure}{.5\textwidth}
    \centering
    \begin{tikzpicture}[scale=1.1]

  \draw[draw=lightgray, rounded corners, thick] (-2.,0.5) rectangle (2,-2.0) {};
  \draw[draw=darkblue!50, rounded corners, thick] (-2.,1.4) rectangle (0,0.6) {};

  \node[circle, scale=0.7, draw=black, fill=lightgray, thick](1) at (-1,0) {\large $A_{ij}$};
 \node[circle, scale=0.7, draw=black, fill=lightgray, thick](2) at (1,0) {\large $A_{ji}$};
 \node[circle,  scale=0.8, draw=darkblue!50, fill=white, thick](3) at (-1.5,-1) {\large $u_{i}$};
 \node[circle,  scale=0.8, draw=darkblue!50, fill=white, thick](4) at (-0.5,-1) {\large $v_{j}$};
 \node[circle,  scale=0.8, draw=darkblue!50, fill=white, thick](5) at (.5,-1) {\large $u_j$};
 \node[circle,  scale=0.8, draw=darkblue!50, fill=white, thick](6) at (1.5,-1) {\large $v_{i}$}; 
 \node[circle,  scale=0.8, draw=darkblue!50, fill=white, thick](7) at (-1.5,1) {\large $\eta$};
 \node[circle,  scale=0.8, draw=darkblue!50, fill=white, thick](8) at (-.5,1){\large $w$};
 \node[circle,  scale=0.8, draw=gray, fill=white, thick](9) at (0.7,1){\large $\mu$};
 \node[circle,  scale=0.8, draw=deg3, fill=white, thick](10) at (1.7,1) {\large $\pi$};
 \node[rotate=0, color = darkblue!50, rotate=90] at (-2.5, 0.7) {\footnotesize \sf $\sigma_{(ij)}=0$};
 \node[rotate=0, color = deg3, rotate=90] at (2.5, 0.7) {\footnotesize \sf $\sigma_{(ij)}=1$};

  \node[rotate=0, color = black] at (0, -1.8) {\footnotesize \sf $\forall(i,j) \in E$};

  \draw [draw=black, <->,  >=latex, thick] (1) -- (2) {};
\draw [draw=darkblue!50, ->,  >=latex, thick] (3) -- (1) {};
 \draw [draw=darkblue!50, ->,  >=latex, thick] (4) -- (1) {};
 \draw [draw=darkblue!50, ->,  >=latex, thick] (5) -- (2) {};
 \draw [draw=darkblue!50, ->,  >=latex, thick] (6) -- (2) {};
 \draw [draw=darkblue!50, ->,  >=latex, thick] (7) -- (1) {};
 \draw [draw=darkblue!50, ->,  >=latex, thick] (7) -- (2) {};
 \draw [draw=darkblue!50, ->,  >=latex, thick] (8) -- (1) {};
 \draw [draw=darkblue!50, ->,  >=latex, thick] (8) -- (1) {};
 \draw [draw=darkblue!50, ->,  >=latex, thick] (8) -- (2) {};
 \draw [draw=deg3, ->,  >=latex, thick] (10) -- (1) {};
 \draw [draw=deg3, ->,  >=latex, thick] (10) -- (2) {};
 \draw [draw=deg3, ->,  >=latex, thick] (9) -- (10) {};
   \draw [draw=darkblue!50, ->,  >=latex, thick] (9) -- (0, 1) {};
    
   \end{tikzpicture}
    \caption[b]{Graphical model representation.}
    \label{fig:relations}
  \end{subfigure}
  \begin{subfigure}{.5\textwidth}
    \centering
    \begin{adjustbox}{width=0.4\linewidth}
    \begin{tikzpicture}[rotate=50, scale=1.6]

      \begin{scope}[every node/.style={color=black,fill=darkblue!50, circle,
                                      inner sep=6pt}]

        \node (a) at (-.5,4.8) {};
        \node (d) at (1.8,4.2) {};
        \node (c) at (-.3,2.8) {};
        \node (e) at (-2,1) {};
        \node (g) at (2,1.5) {};
      \end{scope}

      \begin{scope}[every node/.style={color=black,fill=darkblue!50, circle,
                                      inner sep=6pt}]

        \node (b) at (-2.5,3) {};
        \node (f) at (-0.8,0.8) {};
        \node (h) at (-.5,-1.5) {};
        \node (i) at (1.3,-.0) {};

      \end{scope}

      \begin{scope}[every edge/.style={draw=darkblue!50,very thick}]
        \path[draw=darkblue!50,->,  >=latex, shorten <= 2pt, thick](a)  edge [bend left=15] node [swap] {} (c);
        \path[draw=darkblue!50,->,  >=latex, shorten <= 2pt, thick](c)  edge [bend left=15] node [swap] {} (e);
        \path[draw=deg2,->,  >=latex, shorten <= 2pt, thick](e)  edge [bend left=15] node [swap] {} (c);
        \path[draw=darkblue!50,->,  >=latex, shorten <= 2pt, thick](f)  edge [bend left=-15] node [swap] {} (d);
        \path[draw=darkblue!50,->,  >=latex, shorten <= 2pt, thick](c)  edge [bend left=15] node [swap] {} (d);
        \path[draw=darkblue!50,->,  >=latex, shorten <= 2pt, thick](b)  edge [bend left=-15] node [swap] {} (g);
        \path[draw=darkblue!50,->,  >=latex, shorten <= 2pt, thick](g)  edge [bend left=15] node [swap] {} (c);
        \path[draw=darkblue!50,->,  >=latex, shorten <= 2pt, thick](a)  edge [bend left=-15] node [swap] {} (b);
        \path[draw=darkblue!50,->,  >=latex, shorten <= 2pt, thick](d)  edge [bend left=15] node [swap] {} (c);
        \path[draw=darkorange,->,  >=latex, shorten <= 2pt, thick](d)  edge [bend left=15] node [swap] {} (g);
        \path[draw=darkblue!50,->,  >=latex, shorten <= 2pt, thick](b)  edge [bend left=15] node [swap] {} (c);
        \path[draw=darkblue!50,->,  >=latex, shorten <= 2pt, thick](b)  edge [bend left=-15] node [swap] {} (a);
      \end{scope}

      \begin{scope}[every edge/.style={draw=darkblue!50,very thick}]
        \path[draw=darkblue!50,->,  >=latex, shorten <= 2pt, thick](f)  edge [bend left=-15] node [swap] {} (h);
        \path[draw=darkblue!50,->,  >=latex, shorten <= 2pt, thick](i)  edge [bend left=15] node [swap] {} (h);
        \path[draw=darkblue!50,->,  >=latex, shorten <= 2pt, thick](i)  edge [bend left=-15] node [swap] {} (f);
        \path[draw=darkblue!50,->,  >=latex, shorten <= 2pt, thick](f)  edge [bend left=-15] node [swap] {} (i);
        \path[draw=darkblue!50,->,  >=latex, shorten <= 2pt, thick](h)  edge [bend left=-15] node [swap] {} (f);
      \end{scope}
      \begin{scope}[every edge/.style={draw=darkblue!50, very thick}]
        \path[draw=darkblue!50,->,  >=latex, shorten <= 2pt, thick](c)  edge [bend left=15] node [swap] {} (b);
        \path[draw=darkblue!50,->,  >=latex, shorten <= 2pt, thick](f)  edge [bend left=15] node [swap] {} (c);
      \end{scope}
      \begin{scope}[every edge/.style={draw=deg3, very thick}]
          \path[draw=deg3,->,  >=latex, shorten <= 2pt, thick](g)  edge [bend left=15] node [swap] {} (e);
          \path[draw=deg3,->,  >=latex, shorten <= 2pt, thick] (e)  edge  [bend left=15]  node [swap] {} (g);
      \end{scope}
    \end{tikzpicture}
    \end{adjustbox}
    \vspace{0.2cm}
    \caption[b]{Network example.}
    \label{fig:example}
  \end{subfigure}
  \caption{\textbf{Model visualization}. (a) Graphical model: the entry of the adjacency matrix $A_{ij}$ is determined by the community-related latent variables $u, v, w$ and the reciprocity parameter $\eta$ (blue);  and by the anomaly-related parameters $\pi$ (orange) and the hyper-prior $\mu$ (grey). (b) Example of a possible realization of the model: blue edges display interactions based on  community and reciprocity and the orange ones are anomalous.}  \label{fig:vizexample}
\end{figure}
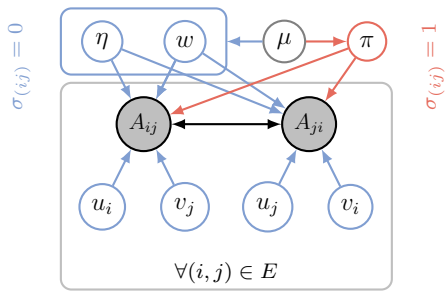
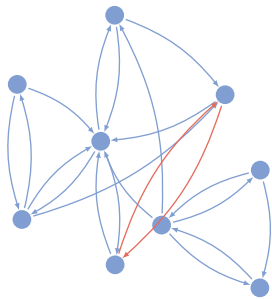 
We aim at modeling reciprocity when two edges are regular, as this can be the result of a reasonable tie formation mechanism involving two nodes, e.g., exchanging favors or cooperative behaviors. For anomalous edges instead, it is less clear what would reciprocity mean, hence we remain agnostic to it and assume that the edges $i\ra j$ and $j\ra i$ are independent when they are anomalous. In other words, the existence of  the anomalous edge $A_{ji}$ has no influence on its reciprocated edge $A_{ij}$, which is also anomalous. To reflect this mathematically, we set $J_{(ij)}^{(a)}=0$. This follows the properties of  multivariate Bernoulli distributions, where independence and uncorrelatedness are equivalent phenomena  \cite{dai2013multivariate}. As the correlation between the pair of edges $(A_{ij},A_{ji})$  is captured by $J_{(ij)}^{(\ell)}$,  when $J_{(ij)}^{(\ell)} = 0$, the pair $(A_{ij},A_{ji})$ is uncorrelated. In addition, we assume a symmetric structure of $f ^{(a)}=f ^{(a)}_{ij}=f ^{(a)}_{ji}$ for all anomalous edges. 

To summarize the steps of  our proposed generative model: we first draw hidden labels for the edges, determining  them being regular or anomalous; then,  we draw pairs of edges $(A_{ij}\,, A_{ji})$ from a specific form of distribution depending on the edges' labels. 
Formally, the generative model is:
\begin{align}
\sigma_{(ij)} &\sim \text{Bern}(\mu) \label{eqn:Zprior2}\\
A_{(ij)} &\sim \begin{cases}  \f{\exp\ccup{(A_{ij}+ A_{ji}) f ^{(a)}}}{Z^{(a)}_{(ij)}} & \text{if} \quad \sigma_{(ij)}=1 \quad   \\ \f{\exp\ccup{A_{ij}f^{(r)}_{ij} + A_{ji}f^{(r)}_{ji} + A_{ij}A_{ji}J^{(r)}_{(ij)}}}{Z^{(r)}_{(ij)}}& \text{if} \quad \sigma_{(ij)} = 0 \quad \end{cases} \label{eqn:Poiss}
\end{align}

Up to this point, we focused on reciprocity and how to incorporate it into our model via the interaction term $J^{(r)}_{(ij)}$. Now, we turn our attention to community structure, another important mechanism that we believe regulates tie formation of regular edges. Conversely, we assume that communities have no influence on anomalous edges. To formalize this, we utilize similar model specifications as outlined in \cite{contisciani2021JointCrep}, and we incorporate  community structure  through latent variables embedded in the natural parameters of the joint pair distribution $P^{(r)}(A_{ij},A_{ji})$.  In detail, we assume the tie formation depends on communities and reciprocity for regular edges, and  only on anomaly parameter for anomalous ties. 
\begin{align}
&f^{(r)}_{ij} = \log \lambda_{ij} \ , \ \label{eqn:fij}  f^{(r)}_{ji}  = \log \lambda_{ji} \ ,\\
&J^{(r)}_{(ij)}  = \log \eta \label{eqn:Jij}  \ ,\\
&f^{(a)}  =  \log \pi \ ,
\end{align}
where 
\be\label{eqn:lambda}
\lambda_{ij} = \sum_{k, q=1}^{K} u_{ik}v_{jq} w_{kq} \quad ,
\ee
regulates how mixed-membership community structure determines tie formation in directed networks, as in \cite{de2017pre}. We provide a  schematic visualization of these contributions in \cref{fig:vizexample}.
The normalization parameters are obtained by enforcing the normalization constraint using the above definitions, so that $Z^{(a)}_{(ij)}  =  (\pi +1)^2$ and $Z^{(r)}_{(ij)}   = \lambda_{ij}+\lambda_{ji}+\eta  \lambda_{ij}\lambda_{ji}+1$.

The parameters $\lambda$ and $\eta$ play important roles in our model of community-reciprocity structure. $\lambda$ captures the mixed-membership aspect, while $\eta$ is the pair-interaction coefficient that regulates the formation of pairs of  edges between nodes. The $K$-dimensional vectors ${u_i}$ and ${v_i}$ represent the out-going and in-coming communities of node $i$, respectively. The entries in these vectors, $u_{ik} \geq 0$ and $v_{jq} \geq 0$, represent the weights assigned to each community, where $K$ is the number of communities. The value of $K$ can be either specified as input or selected using model selection criteria, such as cross-validation \cite{de2017pre}.  The affinity matrix $w_{kq}$ controls the structure of the communities, with higher values on the diagonal indicating more assortative communities.   The formation of anomalous edges is derived by the latent parameter $\pi > 0$, as in the  $\lim {\pi \to 0}$ the probability of the existence of an anomalous edge converges to zero (see \Cref{appendix:derive} for more details on derivations).  All of these parameters, along with $\mu$, are included in the latent parameter set $\boldsymbol \Theta=\{\{u_i\}, \{v_i\},\{w_{kq}\},\eta,\pi,\mu\}$ that will be inferred from data. In addition to  point estimates of these parameters, our model returns a posterior estimate for the edge label variable $\sigma_{(ij)}$ in the form of a Bernoulli posterior distribution of parameter $Q_{(ij)}$. This is also the estimated expected value of the edge label. We provide more details in \cref{sec:inference}. 

Our model assumes that community structure drives the process of  formation of a regular edge, and that the regular edges between a pair of nodes depend on each other explicitly according to the value of $\eta$. If $J^{(r)}_{(ij)}=0$  (when $\eta=1$),  the probability of  the edges  between nodes $i$ and $j$ is determined solely by their respective communities. On the other hand, a positive value  of $J^{(r)}_{(ij)}$ (when $\eta>1$)  increases the probability of  the existence of both $i\ra j$ and $j\ra i$, while a negative value (when $0<\eta<1$)  decreases it. \\

By utilizing properties  of the bivariate Bernoulli distribution \cite{contisciani2021JointCrep,dai2013multivariate}, we  obtain a closed-form  solution for the expected value of an edge (see for more details \Cref{appendix:derive}):
\begin{equation} 
\Exp\rup{A_{ij}} = (1-Q_{(ij)}) \,\f{\lambda_{ij}+\eta \lambda_{ij}\lambda_{ji}}{Z^{(r)}_{(ij)}}+Q_{(ij)} \,\f{\pi}{1+\pi} \label{eqn:marginalmean} \ .
\end{equation} 
This result is useful in link prediction experiments, in that we can score edges based on the values calculated from \cref{eqn:marginalmean} and use these to compute prediction metrics such as the area under the receiver operating curve (AUC), we illustrate this in \Cref{sec:syn_datasets}.

\setlength{\textfloatsep}{5pt}
	\SetKwInOut{Input}{Input} 
\begin{algorithm}[H]
 	\caption{\ajcrep \text{} : {EM algorithm.}}
	\label{alg:EM}
	\SetKwInOut{Input}{Input}
	\setstretch{0.7}
	\raggedright
	\Input{network $\boldsymbol A=\{A_{ij}\}_{i,j=1}^{N}$, \\number of communities $K$.}
  	\BlankLine
	\KwOut{memberships $u=\rup{u_{ik}},\, v=\rup{v_{ik}}$; network affinity matrix $w=\rup{w_{kq}}$; pair-interaction coefficient $\eta$;  anomaly parameter $\pi$;  prior on anomaly indicator $\mu$.}
	\BlankLine
	 Initialize $\boldsymbol \Theta:(u,v,w, \eta,\pi, \mu)$ at random.  
	 \BlankLine
	 Repeat until $L(\boldsymbol \Theta)$ convergence:
	 \BlankLine
	\quad 1. Calculate $\rho$ and $\boldsymbol Q$ (E-step):
		\bea
	 &&\rho_{ijkq}\sim \text{as in Eq.} (\ref{eqn:rho})  \;,\quad \nonumber \\ &&Q_{ij}  \, \sim \text{as in Eq.} (\ref{app:eqn_Qij}) \;.\quad  
	\nonumber
	\eea
	
	 \quad 2. Update parameters $\boldsymbol \Theta$ (M-step):  
	\BlankLine
	\quad \quad \quad  
		i) for each node $i$ and community $k$ update memberships:
		\bea
		\quad u_{ik} = \f{\sum_{jq} (1-Q_{(ij)})\, A_{ij}\rho_{ijkq} }{\sum_{j}\rup{\f{\sum_{q}(1-Q_{(ij)})\,(1+\eta\, \lambda_{ji})\,v_{jq}w_{kq}}{\lambda_{ij}+\lambda_{ji}+\eta  \lambda_{ij}\lambda_{ji}+1}}}  \nonumber\\
	    \quad v_{ik} = \f{\sum_{jq} (1-Q_{(ij)})\, A_{ji}\rho_{jiqk} }{\sum_{j}\rup{\f{\sum_q (1-Q_{(ij)})\,(1+\eta\, \lambda_{ij})\,u_{jq}w_{qk}}{\lambda_{ij}+\lambda_{ji}+\eta  \lambda_{ij}\lambda_{ji}+1}}}  \nonumber
		\eea
	\quad \quad \quad 
	ii) for each pair $(k,q)$ update affinity matrix:
		\be
		\quad w_{kq} = \f{\sum_{i,j} (1-Q_{(ij)})\, A_{ij}\rho_{ijkq} }{\sum_{i,j}\rup{\f{(1-Q_{(ij)})\,(1+\eta\, \lambda_{ij})\, u_{ik}v_{jq}}{\lambda_{ij}+\lambda_{ji}+\eta  \lambda_{ij}\lambda_{ji}+1}}} \nonumber
		 \ee
		 \quad \quad \quad 
		iii) update pair-interaction coefficient:
		\be
		\quad  \eta = \f{\sum_{(i,j)} (1-Q_{(ij)} )\,A_{ij}A_{ji}}{\sum_{(i,j)} (1-Q_{(ij)} )\,\rup{\f{\lambda_{ij}\lambda_{ji}}{\lambda_{ij}+\lambda_{ji}+\eta  \lambda_{ij}\lambda_{ji}+1}}} \nonumber
		 \ee
	\quad \quad \quad 
		iv) update anomaly parameter:
		\be \label{eqn:pi}
		\quad \pi = \f{\sum_{(i,j)}Q_{(ij)}\,(A_{ij}+A_{ji})}{\sum_{(i,j)}Q_{(ij)}\,(2-A_{ij}-A_{ji})} \nonumber
		\ee
		\quad \quad \quad 
		v) update  prior on anomaly indicator:
		\be \label{eqn:mu}
		\quad \mu = \f{1}{N(N-1)/2}\sum_{(i,j)} Q_{(ij)} \nonumber \quad.
		\ee
		\quad \quad \quad 
\end{algorithm} 
\section{\label{sec:inference}Inference}
Our ultimate goal is to determine $\boldsymbol \Theta$, the latent parameters of the model. To do this, we maximize the posterior probability $P(\boldsymbol\Theta|\boldsymbol A) = \sum_{\boldsymbol\sigma}P(\boldsymbol\sigma,\boldsymbol\Theta|\boldsymbol A)$.  Instead of directly maximizing this probability, it is more computationally efficient to maximize the log-posterior, as the maxima of the two functions are equivalent:

\bea 
L(\boldsymbol \Theta) &= \log P(\boldsymbol \Theta| \boldsymbol A) = \log\sum_{\boldsymbol\sigma}P(\boldsymbol \sigma,\boldsymbol \Theta| \boldsymbol A) \nonumber \\\nonumber \\
&\geq  \sum_{\boldsymbol\sigma} q(\boldsymbol \sigma)\, \log \f{P(\boldsymbol \sigma,\boldsymbol \Theta| \boldsymbol A)}{q(\boldsymbol \sigma)} \ ,   \label{eqn:LP} 
\eea
where we defined $q(\boldsymbol \sigma)$, a variational distribution that must sum to 1.
Our maximum likelihood approach  involves the use of an expectation-maximization (EM) algorithm in which we alternately update different sets of parameters of our model. More specifically, we first update the variational distribution  parameters (E-step), $\rho$ and $\boldsymbol Q$, and then maximize $L(\boldsymbol \Theta)$ with respect to $\boldsymbol \Theta$ (M-step). This process is repeated until $L(\boldsymbol \Theta)$ converges, signifying the completion of the optimization process. The full procedure is outlined in \Cref{alg:EM} (see  \Cref{app:inference} for more details on the inference task).
The computational complexity of the algorithm is $O(N^2)$, primarily due to the terms in the dense matrix $Q_{(ij)}$ that are not multiplied by the sparse adjacency  matrix $A_{ij}$.  As $\boldsymbol Q$ is crucial for identifying anomalous edges, its presence may make the model infeasible for large systems. Investigating ways to reduce this complexity, for instance by making its representation sparse, is an interesting avenue for future work.

\section{\label{sec:results}Results} 
\subsection{\label{sec:syn_datasets}Synthetic datasets}  
We validate our model on synthetic datasets, generated with the generative algorithm  in  \cref{appendix:sec_gen}. The studied  datasets consist of  $N=500$ nodes, with an average degree of $\langle k \rangle = 60$. The number of communities is set to $K = 3$, and the pair-interaction coefficient, $\eta$, has a range of values. The anomaly density (ratio of anomalous edges to total number of edges) is varied within the interval $\rho_a\in [0,1]$.   We compare  \ajcrep \text{}  with \jcrep\ \cite{contisciani2021JointCrep}, which is what \ajcrep\ reduces to if we had not considered anomalies, i.e., when $\mu =0$ and $\lim {\pi \to 0}$. This allows to focus on observing the impact of considering the existence of anomalous edges in a given dataset.   

In order to determine the effectiveness of our proposed model, which is based on the concept of community structure, we  first evaluate its ability to accurately identify the memberships of individuals within a community. To accomplish this, we measure the cosine similarity (CS) between the  ground truth and inferred community memberships vectors. The CS has values in $[0,1]$, with  $\text{CS} =1$ indicating the best performance. For this task, we also run a Bayesian Poisson matrix factorization (BPMF) algorithm \cite{gopalan2015scalable}. BPMF is a scalable algorithm for factorizing sparse matrices and provides a useful comparison for our proposed algorithm. We run all algorithms on synthetic datasets generated by \ajcrep \text{} (see  \Cref{appendix:sec_gen} for more details). 
The results, as illustrated in \cref{appendix:fig_syn} (a) and (b), show that when the proportion of anomalous edges in the dataset is relatively low, BPMF outperforms our proposed algorithm. However, when the number of anomalous edges is above $50\%$ of the total number of edges, our algorithm is still able to detect community structure with a reasonable level of accuracy. Additionally, it can be observed that  \ajcrep \text{} performs the same as \jcrep; with both models having higher performance for smaller values of the anomaly density, $\rho_a$. This behavior is expected, as for higher values of $\rho_a$, the community structure plays a weaker role in the formation of edges.

It is worth mentioning that the primary objective of the current research is to develop the capabilities of \jcrep \text{} through the incorporation of anomaly detection functionality, rather than focusing on further improving its community detection abilities or recovering reciprocity parameter. Therefore, our focus is on assessing and optimizing the model's anomaly detection  potential. For this, we measure the AUC on edges,  i.e., on a binary matrix that stores what edges are true anomalies, and use as scores the inferred ${Q}_{(ij)}$.
From our results, illustrated in panels (e) and (f) in \cref{appendix:fig_syn}, we find that \ajcrep \text{} demonstrates good performance in the detection of anomalous edges across a range of anomaly densities. Furthermore, the integration of reciprocity effects is enhancing performance, compared to a model (\ACD) where there is no such effect \cite{SafdariJBD2022}.  
 
In addition to evaluating anomaly detection, we are also interested in assessing the ability of \ajcrep \text{} to identify missing edges, also known as link prediction task. In these experiments we employ a 5-fold cross-validation approach, where the dataset is split in five sets of data.  In each realization, four of these groups are utilized as a training set to infer the parameters $\boldsymbol \Theta$. The remaining group is used as a test set, where the score for each pair  $(A_{ij},A_{ji})$ in the matrix is evaluated to compute the AUC. By iteratively varying which group serves as the test set, we obtain a total of five trials per realization. The final AUC value is determined by averaging the results of these trials. The score of an edge is calculated using the closed-form expression for its marginal probability, as described in \cref{eqn:marginalmean}. As shown in panels (c) and (d) in \cref{appendix:fig_syn}, an increase in the reciprocity parameter results in an increase in the AUC for both \ajcrep \text{} and \jcrep, however, we observe a bigger improvement in terms of AUC of \ajcrep\ over the competitive algorithms. These results indicate that our model becomes more effective in link prediction tasks for higher values of reciprocity.

\subsection{\label{sec:rd_datasets}Real World datasets}
In order to assess the practical utility of our model, we investigate its usage on a variety of real-world data covering applications as food-sharing between bats, social support interactions in a rural community, email exchanges, and online dating. Their sizes range from $N=19$ to $N=3562$, see \Cref{tabSI:data_desc} in \Cref{appendix:data_desc} for a summary description.
\paragraph*{Injecting anomalous edges}
To evaluate the accuracy and precision of the model in detecting anomalous edges, we first need to know the true label of  edges, being anomalous or regular. 
However, one of the challenges in this regard  is the lack of data containing explicit anomalies. 
To address this challenge, we conduct an experiment where we inject $n$ random edges between nodes in a real dataset and label them as anomalous. We  vary $n$ to evaluate the impact of anomaly density $\rho_a=n/E$ on model performance.  We then run our model on this manipulated dataset and infer the expected value $\Exp\rup{\sigma_{(ij)}}=\hat{Q}_{(ij)}\in [0,1]$ for the edge labels, which also indicates the likelihood that the edges between two nodes are anomalous. Based on this, we assign labels to the edges. In this specific experiment, we label the first $n$ pairs $(i,j)$ with the highest values of $\hat{Q}_{(ij)}$ as anomalous edges. We measure the precision as performance metric, this is the fraction of inferred anomalous edges which are correctly classified (in our case--since we fix the number of inferred anomalous to be equal to  the number of injected anomalous edges--this also corresponds to recall, i.e., the fraction of true anomalous edges that are inferred as such).


\begin{figure}[t] 
	\includegraphics[width=0.85\linewidth]{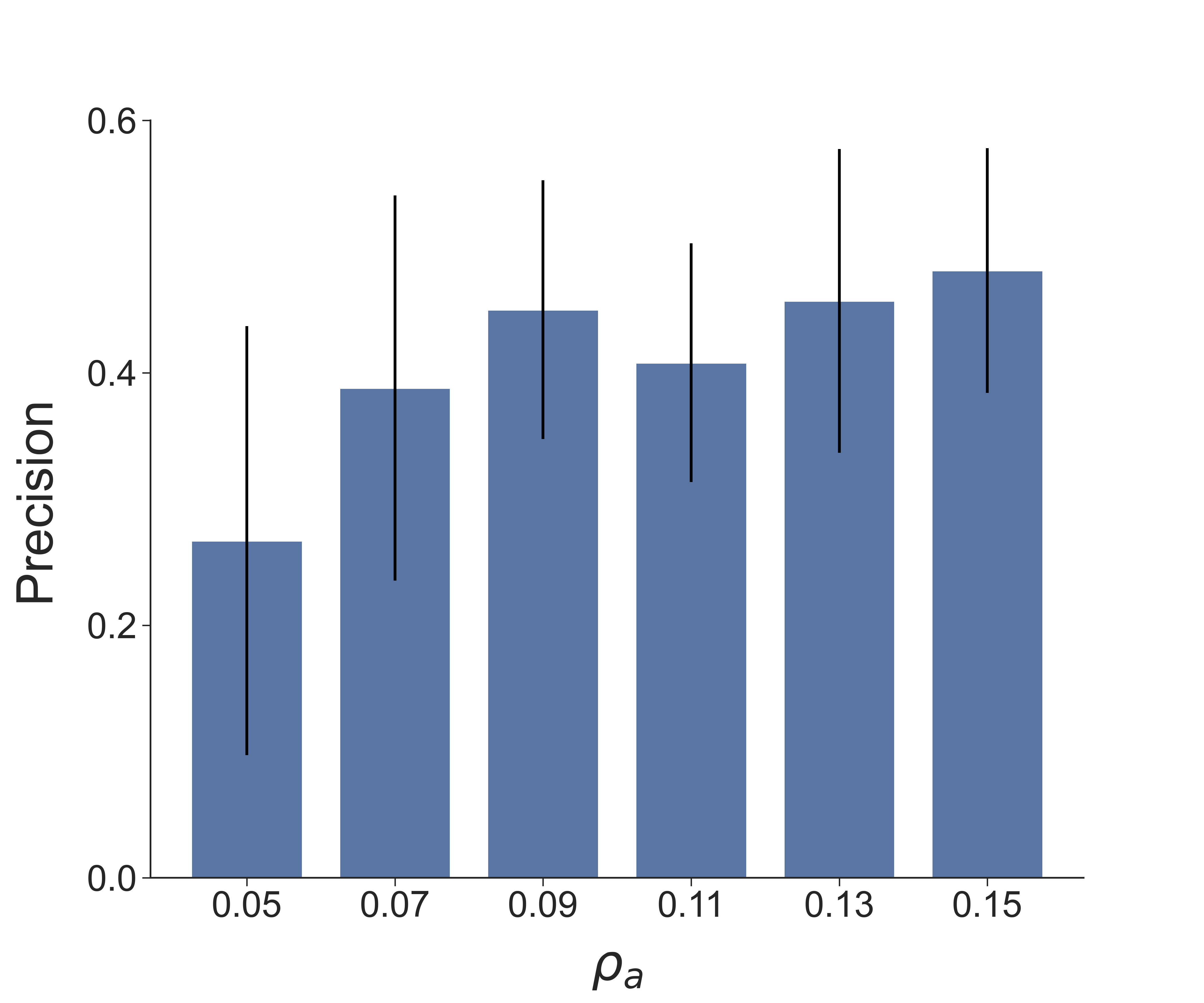}  
	\caption{\textbf{Precision in detecting the  injected edges in the vampire-bat network.}   The precision increases by the increase in the number of anomalous edges injected in the network, i.e., anomaly density in the dataset, $\rho_a$. The results is the average over 10 randomly injected sets of edges, bars are standard deviations.   Here we use the initialization $\pi=0.1$. }
	\label{fig:vampire_precision}
\end{figure}

\subsubsection{Smaller datasets}
\paragraph{Vampire bat network} 
The vampire bat network is a complex and dynamic social structure in which individual vampire bats form connections and share food with one another \cite{carter2013food}. The bats have a remarkable ability to detect the body heat of other bats, even in complete darkness, allowing them to locate potential food sources and potential recipients for food sharing. When a bat finds food, it will often regurgitate some of it and share it with other members of its network. This behavior, known as reciprocal altruism, is essential for the survival of the group, as it ensures that all members have access to food even when they are unable to find it themselves.
The decision of who to feed is likely to be influenced by both the genetic relatedness of the individuals involved and their history of reciprocal sharing. Given this, we expect that reciprocity will play a significant role in determining which individuals form close social ties within this network. As such, when examining this dataset, it will be important to carefully consider this effect. In our analysis, we use the data obtained from \cite{carter2013food} and remove isolated nodes. The network consists of $N=19$ nodes, $E=103$ edges and has high reciprocity of 0.64. In addition, we fix $K=2$ as in \cite{contisciani2021JointCrep}.
\begin{figure}[t] 
	\includegraphics[width=0.7\linewidth]{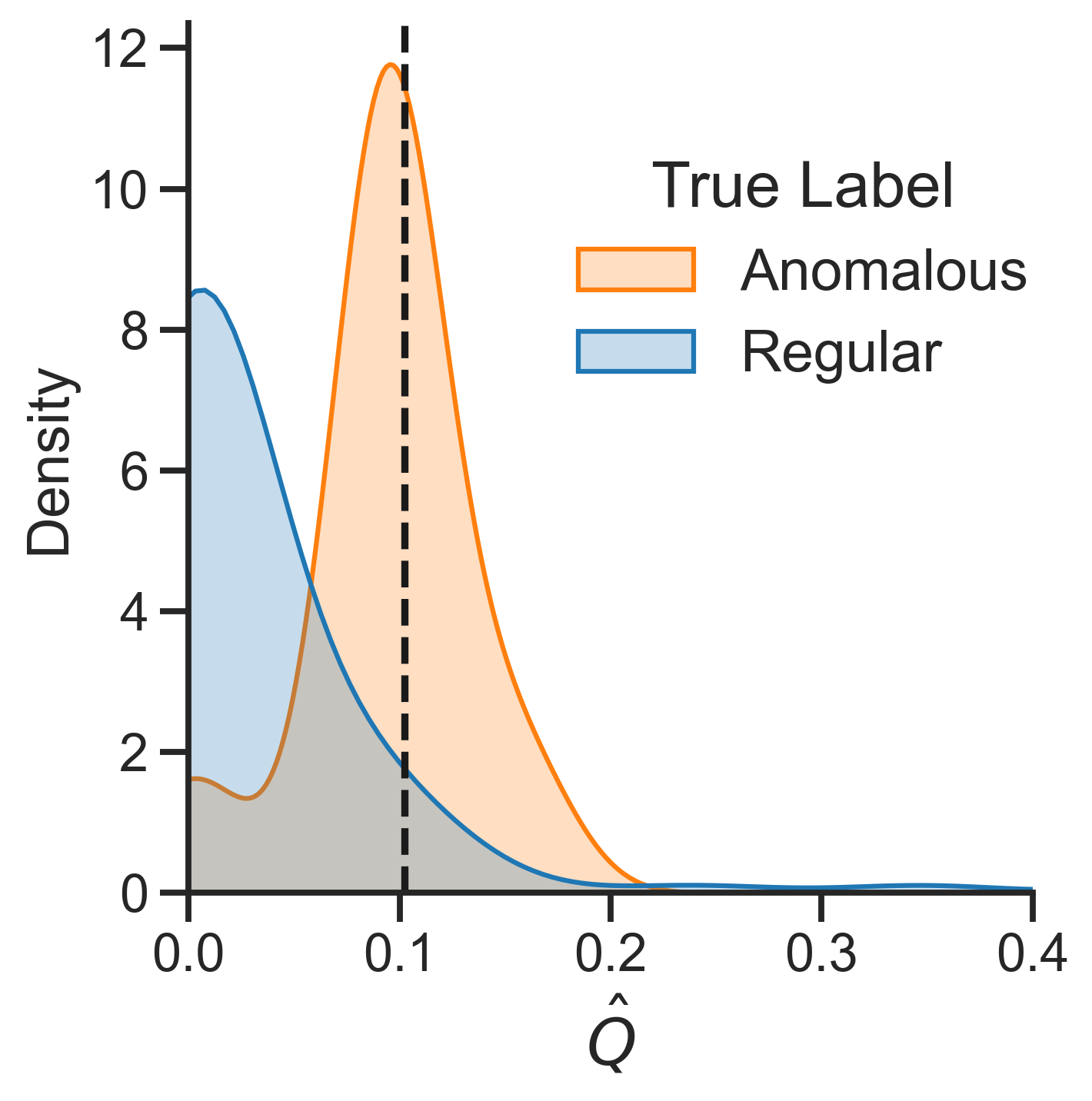}  
	\caption{\textbf{Anomaly Detection in the vampire-bat network.} We show the distribution of $\hat{Q}_{(ij)}$, i.e., the probability of  a pair of edges $(i,j)$ being anomalous, as estimated by \ajcrep. We distinguish the  true  regular and anomalous edges with different colours, blue and orange, respectively, to highlight their different inferred distributions. Here, $\rho_a=0.09$ and $\pi=0.1$. We measure a precision of 0.4. For this, we label as anomalous the fraction of $\rho_a$ edges with highest $\hat{Q}_{(ij)}$. The vertical dashed line denotes the minimum $\hat{Q}_{(ij)}$ observed in this set of anomalous edges.} 
	\label{fig:vampire_QTL}
\end{figure}

As shown in \cref{fig:vampire_precision}, our model's ability to detect anomalies improves when there is a higher concentration of anomalies in the dataset.  The plot depicts the precision in detecting the anomalous edges, for a range of anomaly density, $\rho_{a}$. 
In a more specific case, \cref{fig:vampire_QTL} provides an example of how  \ajcrep \text{}   can be used for anomaly detection in the vampire-bat dataset. In this example, a set of edges with $\rho_a=0.09$ were embedded in the system. In this figure, the entries of the estimated $\hat{Q}$ matrix, which represent the probability of edges being anomalous, are categorized based on their true labels and assigned different colours to highlight their different inferred distributions of $\hat{Q}$. The plot clearly shows two different distributions, one highly picked at $\hat{Q}_{(ij)}=0$ and the other picked around $\hat{Q}_{(ij)}=0.1$. The first corresponds to regular edges, which are thus correctly identified as such, while the latter are the injected anomalies, which are indeed assigned a higher probability of being anomalous. While there are few regular edges that have a high $\hat{Q}$, we observe that a significant density of anomalous edges is concentrated at  $\hat{Q}_{(ij)}>0.1$, indicating that the model is correctly assigning them as anomalous. Quantitatively, we measure precision and recall values of $0.4$, obtained by labelling as anomalous the fraction of $\rho_{a}=0.09$ edges with highest $\hat{Q}_{(ij)}$. Even though a small fraction of regular edges are classified as anomalous and vice versa, these numbers show that overall the algorithm is doing well at detecting the injected anomalies.

\paragraph{A Nicaraguan community}   
The next dataset represents the  social support  network of  indigenous  Nicaraguan  horticulturalists \cite{Jeremy2018}. The original dataset is self-reported  network data. Ties are reported by several individuals and these may be in disagreement with each other. Hence, we process it using \vimure \text{} algorithm \cite{deBacco2021vimure}, which estimates probabilistically an underlying network structure from self-reported network data, provided by multiple reporters, accounting for reciprocity.  The summary description of the estimated network by \vimure \text{} can be seen in \Cref{tabSI:data_desc}.   In addition, it estimates the reliability $\theta>0$ of each individual reporter, with higher values denoting over-reporting. Reliabilities can be correlated to anomalies in that we expect that unreliable reporters may report non-existence ties which we interpret  as anomaly.

To assess this, we run \vimure\ twice. The first time, we run its default version and use it to collect estimates of reporters' reliabilities. The second time, we run it in a modified version where we fix the reliability parameters to a neutral value, assuming that all reporters are reliable. We use this output, the estimated network in this modified version, as input for \ajcrep.  In this way, we aim at observing proxies for anomalous edges: these are some of the edges that are involving unreliable reporters, as estimated in the first run of \vimure. Our model labels anomalies on edges, instead in this dataset we have information on nodes (their reliabilities). We can build a correspondence between these two types of information by assuming that edges connected to the most unreliable reporters would have highest value in the estimated $\hat{Q}$ matrix.  To quantify this match, we assign a value $\hat{Q}_{i}=\max_{j\in\partial i}\hat{Q}_{(ij)}$ to each reporter $i$, where $\partial j$ is its neighbourhood, being the maximum probability that one of its connecting ties is anomalous.

\begin{figure}[t] 
	\includegraphics[width=0.8\linewidth]{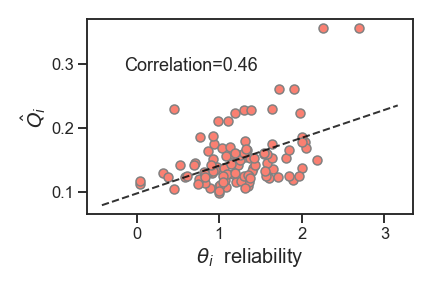}  
	\caption{\textbf{Anomaly Detection in a Nicaraguan social support network.} We show a scatterplot of $\hat{Q}_{i}$ (the maximum probability that one of the connecting ties of node $i$ is anomalous), as estimated by \ajcrep, against $\theta_{i}$, reporters' reliabilities, as estimated by \vimure\ algorithm. The correlation is calculated as the Pearson coefficient, the dashed line is a linear fit to the data. Positive correlation signals that nodes that are more unreliable (high $\theta_{i}$) tend to have an edge that is more likely to be labeled as anomalous among its connections.}
	\label{fig:nicaragua}
\end{figure}

We expect $\hat{Q}_{i}$ to be high for nodes that have a high unreliability $\theta$. We find indeed a positive correlation of 0.46 between $\theta_{i}$ and $\hat{Q}_{i}$, as shown in \cref{fig:nicaragua}. In particular, we observe that the edge $(76,3)$ between the two most unreliable nodes has the maximum observed value of $\hat{Q}_{(ij)}=0.36$, which is consistent with the findings reported in \cite{deBacco2021vimure}. Notice that we expect this correlation to further increase if we were  able  to account explicitly for anomalies on nodes (instead of on edges). In this case, one could envision adapting our formalism to assign random variables $\sigma_{i}$ to nodes, which may result in less tractable distributions and thus higher complexity, but may be more appropriate for applications in which nodes act consistently as anomalous. We leave this as an open question for future work.

\subsubsection{Larger datasets} 
In this section, we test our algorithm on UC Irvine and POK messages, as examples of larger datasets. In each case, we randomly select and add $10\%$ additional edges, labeled as anomalous. The \ajcrep\ algorithm consistently produces reliable results in detecting anomalies in both datasets.

\begin{figure*}[htbp] 
	\includegraphics[width=1\linewidth]{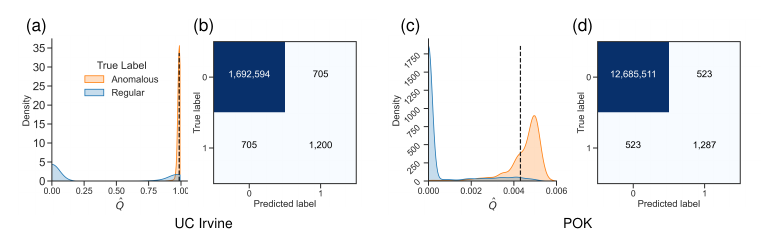}  
	\caption{\textbf{Anomaly Detection in the UC Irvine and POK networks.}  We show the distribution of $\hat{Q}_{(ij)}$, i.e., the probability of  a pair of edges $(i,j)$ being anomalous ((a),(c)) and the confusion matrix ((b),(d)), as estimated by \ajcrep, for the UC Irvine (left) and POK (right) datasets. We distinguish the  true  regular and anomalous edges with different colours, blue and orange, respectively, to highlight their different inferred distributions. Here, $\rho_a=0.1$ and $\pi=0.3$. We measure a precision of 0.63 for UC Irvine and of 0.71 for POK network. The vertical dashed line denotes the minimum $\hat{Q}_{(ij)}$ observed in this set of anomalous edges.}
	\label{fig:ucsocial_pok}
\end{figure*}

\paragraph{UC Irvine messages} 
The network of UC Irvine messages is composed of messages sent between users of an online community of students from the University of California, Irvine \cite{KONECT}. Each node in this communication  network represents a user and each directed edge represents a message that was sent from one user to another.   
Our model consistently identified anomalies in this dataset with a high level of accuracy, as evidenced by  a particularly high peak in the distribution of $\hat{Q}_{(ij)}$ corresponding to anomalous edges in \cref{fig:ucsocial_pok} (a). This is also quantified with a precision value of $0.63$ in the confusion matrix shown in \cref{fig:ucsocial_pok} (b).

\paragraph{Network of online dating}
The POK dataset is a large data set containing the messages exchanged by users within the online dating  POK community.   
The results depicted in \cref{fig:ucsocial_pok} (c-d) demonstrate the strong performance  in identifying and reconstructing  anomalous edges. Figure \ref{fig:ucsocial_pok} (c) illustrates how, also in this case, the distribution of $\hat{ Q}$ values for the anomalous edges is peaked around higher values. While this distribution is more distributed (i.e., has higher variance) than the analogous one observed for UC Irvine, here we observe the distribution corresponding to regular edges being more peaked around zero. This means that in this case the model is distinguishing more clearly the regular edges, with the consequence of obtaining a higher precision of 0.71, as shown in the confusion matrix of \cref{fig:ucsocial_pok} (d).\\
Taken together, these results support the efficacy of our classification methodology.

\section{Conclusion} 
We introduce an expressive generative model to detect edge anomalies in networks that takes into account community membership and reciprocity as main mechanisms driving tie formation. By leveraging these two effects, it is able to detect what edges deviate from a regular behavior and estimate their probability of being anomalous. This inference is performed in a joint learning of  edge  anomalies and mixed-memberships of nodes in communities, thus allowing practitioners to flag potential irregular edges while providing an interpretable community structure. 

In contrast to  common models for anomaly detection that rely on metadata on edges or nodes, our model takes as input only the adjacency matrix and estimates anomaly labels on the edges. It is an unsupervised model, meaning  it does not require any input label to train it.  These features make it particularly relevant in cases where extra information is not available --which is the case for many networked datasets-- where the applicability of many machine learning methods for anomaly detection is significantly limited. As an example, traditional models  for anomaly detection in financial transactions often rely on metadata such as transaction amount, location, and merchant information \cite{Lucas2019,VANVLASSELAER201538,Aleskerov1997}. Instead, our model only requires the adjacency matrix of the transactions, which represents the connections between different account holders. 

One key feature of our model is that it provides a joint probability for the existing pairs of edges between any pairs of nodes, allowing for the inclusion of reciprocity in the model, a relevant property in many directed networks. Furthermore, our model allows for mixed community membership, meaning that nodes can belong to more than one community. This is a more realistic representation of data structures compared to models that assume a single community membership for each node.

There are numbers of ways that our model could be further improved. As mentioned above, our model takes little information in input, only the network's adjacency matrix. A natural next step would be to extend the current model to account for extra information as node attributes, using ideas from generative models with both communities and attributes \cite{newman2016structure,contisciani2020community}; or to consider techniques from semi-supervised learning \cite{zhu2009introduction}, in case of availability of labels on a subset of the edges.\\
Furthermore, we can envision that, for rich and large datasets, deep learning architectures for anomaly detection \cite{liu2022pygod, liu2022bond,Mueller2013} may be competitive methods. However, one could imagine extending standard architectures by combining them with the main ingredients of our model, in datasets where communities and reciprocity matter. The robust performance in detecting anomalies in real data with no extra information suggests that by combining these insights with complex deep architectures may make the latter more expressive and thus boost predictive power. \\
Another type of extra information that is present in many real datasets is time \cite{masuda2016guide}. Edges can be timestamped and this could be used to improve estimates of anomalies. Hence, future work could be directed at generalizing our model to dynamical networks, for instance by combining insights from generative models for dynamic networks with communities \cite{zhang2017random,safdari2022reciprocity,peixoto2017modelling}.

It is important to note that the inferred labels for edges in our model should be treated as estimates rather than definitive conclusions. These labels should be used with caution in the study of a network, as further investigation may be necessary to fully understand the nature of anomalous edges. However, our model can provide valuable insights for practitioners to better understand and interpret the networks they are studying, especially when combined with their specialized knowledge and understanding of the data at hand.

\begin{acknowledgments}
All the authors were supported by the Cyber Valley Research Fund.
The authors thank the International Max Planck Research School for Intelligent Systems (IMPRS-IS) for supporting Martina Contisciani.
\end{acknowledgments}


\clearpage
\newpage

\begin{widetext} 
\appendix

\section{\label{appendix:derive}DETAILED DERIVATIONS} 
\paragraph*{Anomalous edges:} As in the formation of anomalous edges, the reciprocated edges are independent, we apply the condition $J^{(a)}_{(ij)} = 0$, therefore from Eq.(\ref{eqn:ftot}), we find 
\be \label{appendix:Ja0}
\f{p^{(a)}_{11}p^{(a)}_{00}}{p^{(a)}_{10}p^{(a)}_{01}}  = 1\quad \Rightarrow \quad p^{(a)}_{11} =\f{p^{(a)}_{10}p^{(a)}_{01}}{p^{(a)}_{00}} \ .
\ee
Moreover, $f^{(a)}_{(ij)}=f^{(a)}_{(ji)}=f^{(a)} \implies p^{(a)}_{10} = p^{(a)}_{01} = p^{(a)}$ and 
\be\label{appendix:fapi}
f^{(a)} = \log \pi = \log \f{p^{(a)}}{p^{(a)}_{00}} \quad \Rightarrow \quad p^{(a)} = \pi \, p^{(a)}_{00} \ .
\ee
Using the normalization condition, $p^{(a)}_{00}+p^{(a)}_{10}+p^{(a)}_{01}+p^{(a)}_{11}=1$, and the results of \Crefrange{appendix:Ja0}{appendix:fapi}, we find the explicit mapping between the latent variables and the instances of $P^{(a)}(A_{ij}, A_{ji}|\theta_a)$ in \Cref{eqn:P_joint},  
\be
p^{(a)}_{00} =\f{1}{Z^{(a)}_{ij}}, \quad p^{(a)}_{10} = p^{(a)}_{01} =\f{\pi}{Z^{(a)}_{ij}}, \quad p^{(a)}_{11} =\f{\pi^2}{Z^{(a)}_{ij}} \ , 
\ee
where the normalization constant is:
\be\label{appendix:eqnZa}
Z^{(a)}_{ij} = (1+\pi)^2 \ .
\ee
\paragraph*{Regular  edges:} In order to find the explicit mapping between the latent variables and the instances of $P^{(r)}(A_{ij}, A_{ji}|\theta_r)$ in \Cref{eqn:P_joint}, we follow the same procedure as in \cite{contisciani2021JointCrep}, 
\begin{align}
p^{(r)}_{01} &= \f{\lambda_{ji}}{Z^{(r)}_{(ij)}} \label{appendix:eqnp01} \\
p^{(r)}_{10} &= \f{\lambda_{ij}}{Z^{(r)}_{(ij)}} \label{appendix:eqnp10} \\
p^{(r)}_{11} &= \f{\eta\lambda_{ij}\lambda_{ji}}{Z^{(r)}_{(ij)}} \label{appendix:eqnp11} \\
p^{(r)}_{00} &= \f{1}{Z^{(r)}_{(ij)}} \label{appendix:eqnp00} \ ,
\end{align}
where the normalization constant is:
\be\label{appendixeqn:Z}
Z^{(r)}_{(ij)} = \lambda_{ij}+\lambda_{ji}+\eta  \lambda_{ij}\lambda_{ji}+1 \ . 
\ee
\paragraph*{} Having these mappings, we can construct the marginal and conditional distributions of the ties. 
Thus, the marginal and conditional distributions of $A_{ij}$  have the following densities, respectively:
\begin{align}
P(A_{ij}) =\rup{[p^{(r)}_{10}]^{A_{ij}}[p^{(r)}_{00}]^{(1-A_{ij})})+ [p^{(r)}_{11}]^{A_{ij}} [p^{(r)}_{01}] ^{(1-A_{ij})}}  \times (1-\mu) +\rup{[p^{(a)}_{10}]^{A_{ij}}[p^{(a)}_{00}]^{(1-A_{ij})}+ [p^{(a)}_{11}]^{A_{ij}} [p^{(a)}_{01}] ^{(1-A_{ij})}}  \times \mu \ ,
\label{eqn:marginals} 
\end{align}


\begin{align}
P(A_{ij}| A_{ji}) &=  \f{[p^{(r)}_{11}]^{A_{ij}\,A_{ji}} [p^{(r)}_{10}]^{A_{ij}\,(1-A_{ji})} [p^{(r)}_{01}]^{(1-A_{ij})\,A_{ji}}  [p^{(r)}_{00}]^{(1-A_{ij})\,(1-A_{ji})}}{P(A_{ji})} \times (1-\mu)  \nonumber \\\nonumber \\
 &+\f{ [p^{(a)}_{11}]^{A_{ij}\,A_{ji}} [p^{(a)}_{10}]^{A_{ij}\,(1-A_{ji})} [p^{(a)}_{01}]^{(1-A_{ij})\,A_{ji}}  [p^{(a)}_{00}]^{(1-A_{ij})\,(1-A_{ji})}}{P(A_{ji})}   \times \mu  \nonumber \\ 
& \label{eqn:conditionals}
\end{align}

\section{\label{app:inference}INFERENCE}
Our goal is, given  two mechanisms responsible for edge formation, first to determine the values of the  parameters $\boldsymbol \Theta=\{\{u_{ik}\}, \{v_{ik}\},\{w_{kq}\},\eta,\pi, \mu\}$, which  determine  the  relationship between the anomaly indicator $\sigma_{(ij)}$ and the data, and then, given those values, to estimate the indicator $\sigma_{(ij)}$ itself.

We have the posterior:
\be\label{eqn:posterior}
P(\boldsymbol  \sigma,\boldsymbol \Theta| \boldsymbol A) = \f{P(\boldsymbol A|\boldsymbol \sigma,\boldsymbol \Theta) P(\boldsymbol  \sigma|\mu) P(\boldsymbol \Theta)P(\mu)}{P(\boldsymbol A)} \ .
\ee
Summing over all the possible indicators we have:
\be
P(\boldsymbol \Theta| \boldsymbol A) = \sum_{\boldsymbol \sigma}P(\boldsymbol \sigma,\boldsymbol \Theta| \boldsymbol A) \ ,
\ee 
which is the quantity that we need to maximize to extract the optimal $\Theta$. 
It is more convenient to maximize its logarithm, log-posterior, as the two maxima coincide. We use the Jensen's inequality:
\be\label{eqn:jensen} 
L(\boldsymbol \Theta) = \log P(\boldsymbol \Theta| \boldsymbol A) = \log\sum_{\boldsymbol \sigma}P(\boldsymbol \sigma,\boldsymbol \Theta|\boldsymbol A) \geq  \sum_{\boldsymbol \sigma} q(\boldsymbol \sigma)\, \log \f{P(\boldsymbol \sigma,\boldsymbol \Theta| \boldsymbol A)}{q(\boldsymbol \sigma)} \ ,   
\ee
where $q(\boldsymbol \sigma)$ is a variational distribution that must sum to $1$. In fact, the exact equality happens when,
\be\label{eqn:q} 
q(\boldsymbol \sigma) = \f{P(\boldsymbol \sigma,\boldsymbol \Theta|\boldsymbol A)}{\sum_{\boldsymbol\sigma } P(\boldsymbol \sigma,\boldsymbol \Theta|\boldsymbol A)}\ ,
\ee 
this definition is also equivalent to maximizing the right-hand-side of Eq.~(\ref{eqn:jensen}) w.r.t. $q$. 

Finally, we  need to maximize the  log-posterior with respect to $\boldsymbol \Theta$ to get the latent variables. This can be done in an iterative way using Expectation-Maximization algorithm (EM), alternating between maximizing w.r.t. $q$  using  Eq.~(\ref{eqn:q}) and then maximazing Eq.~(\ref{eq:Lconv}) w.r.t. $\boldsymbol \Theta$. In this work, we only fix priors for the $\sigma_{ij}$ (Bernoulli distributions with parameter $\mu$). For this variable we can thus estimate full posterior distributions; instead for the other parameters our model outputs point estimates. This could be modified by suitably specifying priors also for the reciprocity or community-related parameters. In this case, one could easily obtain maximum a posteriori (MAP) estimates with calculations similar to those reported here.\\

Defining $Q_{(ij)} = \sum_{\sigma_{(ij)}} q(\sigma_{(ij)})\, \sigma_{(ij)}$, the expected value of $\sigma_{(ij)}$ over the variational distribution, we obtain,\\
\bea
L(\boldsymbol \Theta) &&=  -\sum_{\boldsymbol \sigma} \, \rup{q(\boldsymbol \sigma) \log  q(\boldsymbol \sigma) } +  \sum_{(i,j)}  \left\{  (1-Q_{(ij)} ) \left( A_{ij}\, f^{(r)}_{ij}+A_{ji}\, f^{(r)}_{ji}+A_{ij}A_{ji}\, J^{(r)}_{(ij)}- \log Z^{(r)}_{(ij)} \right) +   \right.  \nonumber \\ \nonumber \\
&& \left.  +\, Q_{(ij)}  \Big( (A_{ij}+A_{ji})\,f^{(a)}-\log Z^{(a)}_{(ij)}  \Big) + Q_{(ij)}  \log \mu + (1-Q_{(ij)} ) \log (1-\mu) \right\}  \quad,   
\eea
and having Eqs.~(\ref{eqn:fij}-\ref{eqn:lambda}), 
\bea
L (\boldsymbol \Theta)&&=  -\sum_{\boldsymbol \sigma} \, \rup{q(\boldsymbol \sigma) \log  q(\boldsymbol \sigma) } + \\
&& + \, \sum_{(i,j)}  \left\{  (1-Q_{(ij)} ) \Bigg( A_{ij}\, \log\sum_{k} u_{ik}v_{jq}w_{kq}+A_{ji}\, \log\sum_{k} u_{jk}v_{iq}w_{kq}+ A_{ij}A_{ji}\, \log \eta  \right.     \nonumber \\
&& \left. - \log \Big[ \sum_{k,q} u_{ik}v_{jq}w_{kq} + \sum_{k,q} u_{jk}v_{iq}w_{kq} +\eta \sum_{k,q}u_{ik}v_{jq}w_{kq}\sum_{k,q}u_{jk}v_{iq}w_{kq} +1 \Big] \Bigg)   \right.  \nonumber \\
&& \left.  \right.  \nonumber \\
&& \left. + Q_{(ij)}  \Big( (A_{ij}+A_{ji} ) \, \log \pi -2\, \log (\pi +1)  \Big)  +  Q_{(ij)}  \log \mu +   (1-Q_{(ij)}) \log (1-\mu) \right\}  \ .
\eea

Derivative of  log-posterior w.r.t $\eta$,
\bea \label{eqn:dereta}
\f{\partial L (\boldsymbol \Theta)}{\partial \eta} =  \f{1}{ \eta}\sum_{(i,j)}(1-Q_{(ij)} )\,A_{ij}A_{ji} -  \sum_{(i,j)}(1-Q_{(ij)} )\,\f{\lambda_{ij}\lambda_{ji}}{\lambda_{ij}+\lambda_{ji}+\eta  \lambda_{ij}\lambda_{ji}+1}\overset{!}{=}0 
\eea
leads to a fixed-point equation,
\be  \label{eqn:eta}
\eta = f(\eta)= \f{\sum_{(i,j)} (1-Q_{(ij)} )\,A_{ij}A_{ji}}{\sum_{(i,j)} (1-Q_{(ij)} )\,\rup{\f{\lambda_{ij}\lambda_{ji}}{\lambda_{ij}+\lambda_{ji}+\eta  \lambda_{ij}\lambda_{ji}+1}}} \ ,
\ee
which can be solved numerically with fixed-point methods. Alternatively, one can use root-finding methods to solve directly Eq.~(\ref{eqn:dereta}) in $\eta$.\\
 
The equations for the remaining parameters need to be solved using Jensen's inequality, and using  $\log x < x$ to obtain  $-\log x > -x$    \\
\bea
L (\boldsymbol \Theta) &\geq& -\sum_{\boldsymbol \sigma} \, \rup{q(\boldsymbol \sigma) \log  q(\boldsymbol \sigma) } + \\
&& +\,\sum_{(i,j)}  \left\{  (1-Q_{(ij)} ) \Bigg( A_{ij}\, \sum_{k,q}\rho_{ijkq} \log\Big( \f{u_{ik}v_{jq}w_{kq}}{\rho_{ijkq}}\Big)\,+A_{ji}\, \sum_{k,q}\rho_{jikq} \log\Big( \f{u_{jk}v_{iq}w_{kq}}{\rho_{jikq}}\Big)\,\right.   \\
&& \left. + A_{ij}A_{ji}\, \log \eta - \Big[ \sum_{k,q} u_{ik}v_{jq}w_{kq} + \sum_{k,q} u_{jk}v_{iq}w_{kq} +\, \eta \sum_{k,q}u_{ik}v_{jq}w_{kq}\sum_{k,q}u_{jk}v_{iq}w_{kq} +1  \Big] \Bigg)   \right. \nonumber \\ 
&& \left. \right. \nonumber \\ 
&& \left. + Q_{(ij)}  \Big( (A_{ij}+A_{ji}) \, \log \pi -2 \log (\pi +1)  \Big) +  Q_{(ij)}  \log \mu +   (1-Q_{(ij)} ) \log (1-\mu) \right\}
\eea
and the equality holds when 
\be \label{eqn:rho}
\rho_{ijkq}=\f{u_{ik}v_{jq}w_{kq}}{\sum_{k,q}u_{ik}v_{jq}w_{kq}} \ . 
\ee


We derive community parameters, for example we start  by considering $u_{ik}$
\bea
\f{\partial L(\boldsymbol \Theta)}{\partial u_{ik}}&=& \sum_{j} \rup{ (1-Q_{(ij)}) \Big[  A_{ij} \sum_{q} \rho_{ijkq} \f{1}{u_{ik}} - \sum_{q} v_{jq}w_{kq} -  \sum_{q} \eta \, v_{jq}w_{kq}   \lambda_{ji}\Big] } \overset{!}{=} 0 
\eea
and we finally obtain
\be \label{eqn:u}
u_{ik} = \f{\sum_{jq} (1-Q_{(ij)})\, A_{ij}\rho_{ijkq} }{\sum_{j}\rup{\f{\sum_{q}(1-Q_{(ij)})\,(1+\eta\, \lambda_{ji})\,v_{jq}w_{kq}}{\lambda_{ij}+\lambda_{ji}+\eta  \lambda_{ij}\lambda_{ji}+1}}}  \ .
\ee
We find similar expression for $v_{ik}$ and $w_{kq}$:
\be\label{eqn:v}
v_{ik} = \f{\sum_{jq} (1-Q_{(ij)})\, A_{ji}\rho_{jiqk} }{\sum_{j}\rup{\f{\sum_q (1-Q_{(ij)})\,(1+\eta\, \lambda_{ij})\,u_{jq}w_{qk}}{\lambda_{ij}+\lambda_{ji}+\eta  \lambda_{ij}\lambda_{ji}+1}}}
\ee

\be\label{eqn:w}
w_{kq} = \f{\sum_{i,j} (1-Q_{(ij)})\, A_{ij}\rho_{ijkq} }{\sum_{i,j}\rup{\f{(1-Q_{(ij)})\,(1+\eta\, \lambda_{ij})\, u_{ik}v_{jq}}{\lambda_{ij}+\lambda_{ji}+\eta  \lambda_{ij}\lambda_{ji}+1}}} \ .
\ee
For the $\pi$ it yields to the following:
\be \label{eqn:pi}
\pi = \f{\sum_{(i,j)}Q_{(ij)}\,(A_{ij}+A_{ji})}{\sum_{(i,j)}Q_{(ij)}\,(2-A_{ij}-A_{ji})} \ .
\ee
Similarly for $\mu$:
\bea
\f{\partial L(\boldsymbol \Theta)}{\partial \mu} &=&  \sum_{(i,j)}  \f{1}{\mu}\,Q_{(ij)} - \f{1}{1-\mu} \sum_{(i,j)}\,(1-Q_{(ij)})  \overset{!}{=}0
\eea
yielding
\be \label{eqn:mu}
\mu = \f{1}{N(N-1)/2}\sum_{(i,j)} Q_{(ij)} .
\ee

To evaluate $q(\boldsymbol \sigma)$, we substitute the estimated parameters inside Eq.~(\ref{eqn:q}):
\bea
q(\boldsymbol\sigma) &=& \f{\prod_{(i,j)}\;   \Bigg[ \f{ \exp \ccup{ \big( A_{ij}+A_{ji} \big)\,f^{(a)}  }}{Z^{(a)}_{(ij)}}\times  \mu  \Bigg]^{\sigma_{(ij)}}  \Bigg[ \f{ \exp \ccup{ A_{ij}f^{(r)}_{ij}  + A_{ji}f^{(r)}_{ji}+ A_{ij}A_{ji}\,J^{(r)}_{(ij)}}}{Z^{(r)}_{(ij)}}\times  (1-\mu)  \Bigg]^{1-\sigma_{(ij)}} }
{\sum_{\sigma_{(ij)}}\prod_{(i,j)}\; \Bigg[ \f{ \exp \ccup{ \big( A_{ij}+A_{ji} \big)\,f^{(a)}  }}{Z^{(a)}_{(ij)}}\times  \mu  \Bigg]^{\sigma_{(ij)}}   \Bigg[ \f{ \exp \ccup{ A_{ij}f^{(r)}_{ij}  + A_{ji}f^{(r)}_{ji}+ A_{ij}A_{ji}\,J^{(r)}_{(ij)}}}{Z^{(r)}_{(ij)}}\times  (1-\mu)  \Bigg]^{1-\sigma_{(ij)}}  } \nonumber \\ \nonumber \\
&=& \prod_{(i,j)}\; \f{ \Bigg[ \f{ \exp \ccup{ \big( A_{ij}+A_{ji} \big)\,f^{(a)}  }}{Z^{(a)}_{(ij)}}\times  \mu  \Bigg]^{\sigma_{(ij)}} \, \Bigg[ \f{ \exp \ccup{ A_{ij}f^{(r)}_{ij}  + A_{ji}f^{(r)}_{ji}+ A_{ij}A_{ji}\,J^{(r)}_{(ij)}}}{Z_{(ij)}}\times  (1-\mu)  \Bigg]^{1-\sigma_{(ij)}}}
{\sum_{\sigma_{(ij)}=0,1}  \Bigg[ \f{ \exp \ccup{ \big( A_{ij}+A_{ji} \big)\,f^{(a)}  }}{Z^{(a)}_{(ij)}}\times  \mu  \Bigg]^{\sigma_{(ij)}} \,\Bigg[ \f{ \exp \ccup{ A_{ij}f^{(r)}_{ij}  + A_{ji}f^{(r)}_{ji}+ A_{ij}A_{ji}\,J^{(r)}_{(ij)}}}{Z^{(r)}_{(ij)}}\times  (1-\mu)  \Bigg]^{1-\sigma_{(ij)}}} \nonumber \\ \nonumber \\
&=& \prod_{(i,j)}\, Q_{(ij)}^{\sigma_{(ij)}}\, (1-Q_{(ij)})^{(1-\sigma_{(ij)})} \ , \label{eqn:qQij}
\eea
where
\bea
Q_{(ij)} &=& \f{\exp[(A_{ij}+A_{ji})\,f^{(a)}\,- \log Z^{(a)}_{(ij)}]\, \mu  }{\exp[(A_{ij}+A_{ji})\, f^{(a)}\, - \log Z^{(a)}_{(ij)})]\, \mu  +\exp[f^{(r)}_{ij}\,A_{ij}\,+f^{(r)}_{ji}\,A_{ji}\,+J^{(r)}_{(ij)}\,A_{ij}A_{ji}-\log Z^{(r)}_{(ij)}]\, (1-\mu)  } \quad \nonumber \\ \nonumber \\
&=& \f{\exp[(A_{ij}+A_{ji})\, \log \pi\, - 2\log (\pi+1)]\, \mu  }{\exp[(A_{ij}+A_{ji})\, \log \pi\, - 2\log (\pi+1)]\, \mu  +\exp[A_{ij}\log \lambda_{ij}\,+A_{ji}\,\log \lambda_{ji}\,+ \, \log \eta\, A_{ij}A_{ji}-\log Z^{(r)}_{(ij)}]\, (1-\mu)  } \,  \nonumber \\ \nonumber \\
&=& \f{\f{\pi^{( A_{ij}+A_{ji} )} \,\mu}{Z^{(a)}_{(ij)}}}{ \f{\pi^{( A_{ij}+A_{ji} )} \,\mu}{Z^{(a)}_{(ij)}}+\f{\lambda_{ij} ^{A_{ij}}\, \lambda_{ji} ^{A_{ji}}\, \eta^{A_{ij}\,A_{ji}}\,(1-\mu)}{Z^{(r)}_{(ij)}} } \ . \label{app:eqn_Qij}
\eea 
Notice  that this is exactly the expected value w.r.t. the variational distribution as previously defined.\\
 
\subsection{\label{app:conv_crit}Convergence criteria}
The EM algorithm consists of randomly initializing  $u,v,w,\eta,\pi, \mu$, then iterating Eqs.~\ref{eqn:rho}, \ref{app:eqn_Qij}, \ref{eqn:u}-\ref{eqn:w},  \ref{eqn:eta}, \ref{eqn:pi}, \ref{eqn:mu}, until the convergence of the following log-posterior, 
\bea \label{eq:Lconv}
L (\boldsymbol \Theta)  &&= \log P(\boldsymbol\Theta|A) \geq {\sum_{\boldsymbol\sigma} q(\boldsymbol\sigma) \log \f{P(\boldsymbol\sigma,\boldsymbol\Theta|A)}{q(\boldsymbol\sigma)}} \nonumber \\
&& = -{\sum_{\boldsymbol\sigma} q(\boldsymbol\sigma) \log q(\boldsymbol\sigma)}+{\sum_{\boldsymbol\sigma} q(\boldsymbol\sigma) \left\{ \log P(A|\boldsymbol\sigma;\boldsymbol \Theta)+ \log P(\boldsymbol\sigma| \mu)  \right\}}  \nonumber  \\
&& =  -{\sum_{\boldsymbol\sigma} q(\boldsymbol\sigma) \log q(\boldsymbol\sigma)} \nonumber \\
&&+ \, \sum_{\sigma_{(ij)}} q(\sigma_{(ij)}) \left\{  \sum_{(ij)} \rup{ (1-\sigma_{(ij)}) \left( A_{ij}\,f^{(r)}_{ij}+A_{ji}\,f^{(r)}_{ji}+ A_{ij}A_{ji}\,J^{(r)}_{(ij)}- \log Z^{(r)}_{(ij)}\right) +\sigma_{(ij)} \left((A_{ij}+A_{ji})\, f^{(a)}- \log Z^{(a)}_{(ij)} \right) \right. \right.\nonumber \\
&& \left. \left. +\sigma_{(ij)} \log \mu +(1-\sigma_{(ij)})\log (1-\mu) }  \right\} \quad  \nonumber \\ 
\nonumber \\
&& =  -\sum_{(i,j)} \rup{Q_{(ij)}\log Q_{(ij)} +(1-Q_{(ij)}) \log (1-Q_{(ij)}) }  \nonumber  \\
&& +\, \sum_{(i,j)}  \left\{  (1-Q_{(ij)} ) \left( f^{(r)}_{ij}\,A_{ij}+A_{ji}\, f^{(r)}_{ji}+ A_{ij}\,A_{ji}\, J^{(r)}_{(ij)}- \log Z^{(r)}_{(ij)} \right) +   \right. \nonumber  \\
&& \left.  + \, Q_{(ij)}  \left( (A_{ij}+A_{ji})\, f^{(a)}- \log Z^{(a)}_{(ij)}  \right) + Q_{(ij)}  \log \mu +  (1-Q_{(ij)} ) \log (1-\mu) \right\} +const \quad,
\eea
where we neglect $const$, constant terms due to the uniform priors. To calculate $ q(\boldsymbol \sigma)$, we used Eq. (\ref{eqn:qQij}), i.e., a Bernoulli distribution.\\
 
\section{\label{appendix:sec_gen}GENERATIVE MODEL} 
Being generative, the model can be used to generate synthetic networks with anomalies. For this, one should sample the latent parameters $\boldsymbol \Theta=(u,v,w,\eta,\pi,\mu)$, then sample $\boldsymbol\sigma$ given the parameters. Finally, given the $\boldsymbol\sigma$ and the latent parameters,  the adjacency matrix $\boldsymbol A$  could be constructed.
For a given set of community parameters as the input parameters, \cite{de2017pre, SafdariJBD2022},  the expected number of anomalous  and non-anomalous edges are $N^{2}\, \mu  \, \f{\pi}{(1+\pi)} $, and $\Exp\rup{M}=(1-\mu) \,\sum_{i,j}\f{\lambda_{ij}+\eta \lambda_{ij}\lambda_{ji}}{Z^{(r)}_{(ij)}}$, respectively. Assuming a desired total number of edges $ E$, we can thus multiply $\pi, \mu$ and $M$ by suitable sparsity constants that tune: i) the ratio of anomalous edges to the total number of edges,  $\rho_{a}=N^{2}\, \mu   \, \f{\pi}{(1+\pi)} /(N^{2}\, \mu  \, \f{\pi}{(1+\pi)} +(1-\mu) \, \Exp\rup{M}) \in \rup{0,1}$; ii) the success rate of anomalous edges $\pi$.  Once these two are fixed, the remaining sparsity parameter for the matrix $M$, is estimated as: 
\bea
 E\,(1-\rho_{a}) &=&  (1-\mu) \, \sum_{i,j}\f{\zeta \, \lambda_{ij}+\eta \, \zeta \, \lambda_{ij}\, \zeta\, \lambda_{ji}}{\zeta\,\lambda_{ij}+\zeta\,\lambda_{ji}+\eta \,\zeta\, \lambda_{ij}\,\zeta\,\lambda_{ji}+1}
\eea
which can be solved with root-finding methods.\\ 
\clearpage

\section{RESULTS ON SYNTHETIC NETWORKS} 
\begin{figure*}[h]  
	\includegraphics[width=0.8\linewidth]{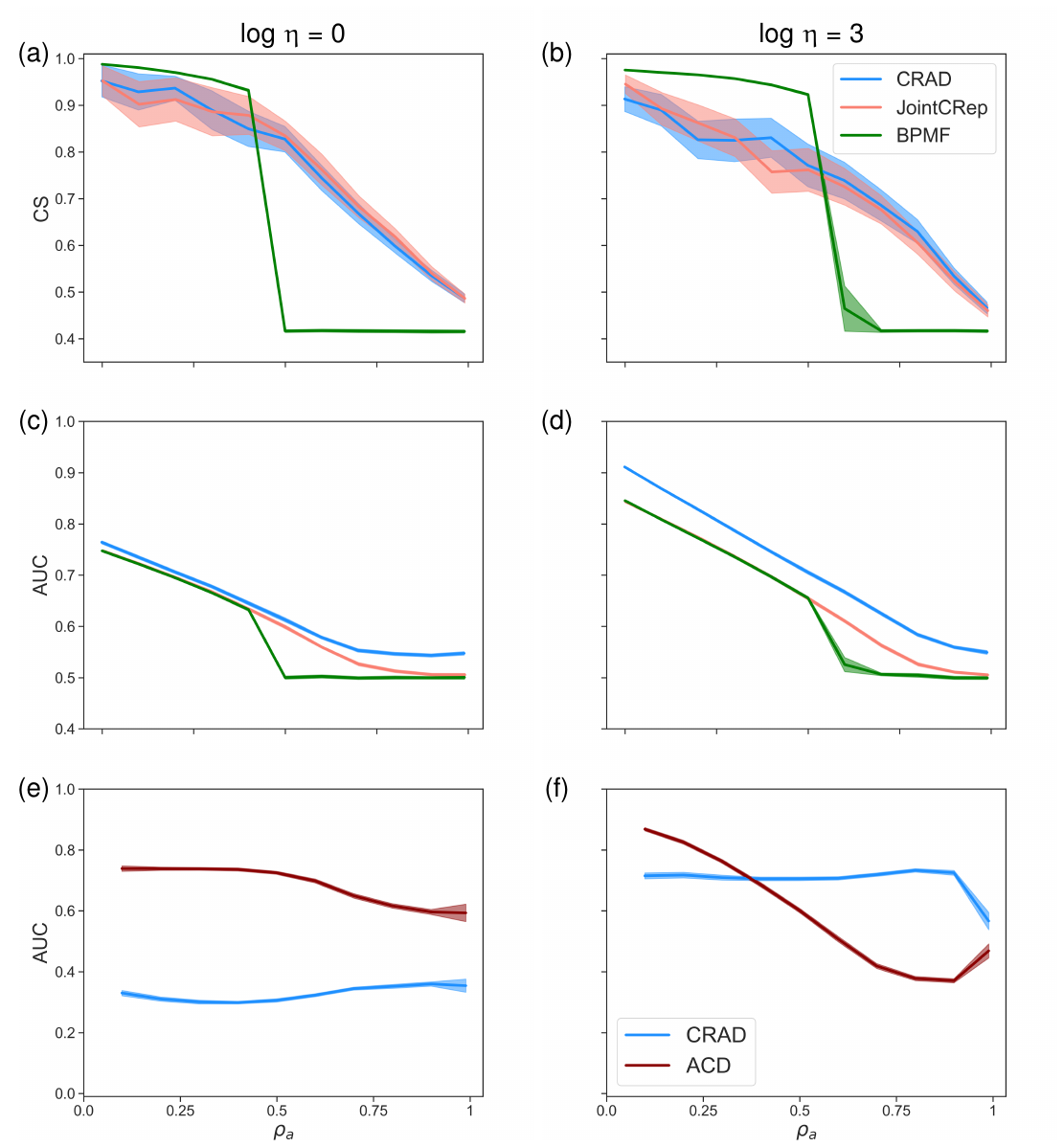} 
	\caption{\textbf{Community detection, link prediction, and anomaly detection on synthetic network datasets.} (a)-(b) We compare the performance of \ajcrep \text{} against JointCrep and BPMF algorithms in community detection, as measured by cosine similarity (CS); and (c)-(d) in link prediction tasks, as measured by AUC on held-out data. In addition, (e)-(f) we test the ability to detect anomalies against a model that does not include a reciprocity effect (\ACD), as measured by the AUC on a binary dataset that contains what edges are regular and what are anomalous. The datasets have $N=500$, average degree $\langle k \rangle=60$, $K=3$.  The first column is for networks generated without reciprocity, $\log \eta = 0$, while the  second column is for networks with positive reciprocity, $\log \eta = 3$. In the $x$-axis we vary $\rho_{a}$, the ratio of anomalous edges over the total number of edges.  Lines and shades around them are averages and standard deviations over 10 network samples, respectively.} 
	\label{appendix:fig_syn}
\end{figure*}
 \clearpage
 
\section{\label{appendix:data_desc} REAL DATA: DATASET DESCRIPTION}

\Cref{tabSI:data_desc} provides a summary of the key characteristics of the studied datasets. The dataset of UC Irvine messages and Online dating  (POK0)  have undergone pre-processing that involved the removal of self-loops, retaining only nodes with both incoming and outgoing edges, and using only the giant connected components.
\begin{table}[h!]
\caption{{\bf {Real-world datasets description.}}} 
\begin{ruledtabular}
\begin{tabular}{lllllll}
 \textbf{Network} & \textbf{Abbreviation}    &$N$ &$E$& \textbf{$\langle k \rangle$} & \textbf{reciprocity} & \textbf{Ref.}\\ \hline
Vampire bat   \hspace{30pt}  & vampire bat   \hspace{10pt}  & $19$   \hspace{10pt}     & $103$     \hspace{10pt}  & $10.8$     \hspace{10pt}  & $0.64$     & \cite{carter2013food}    \\
A Nicaraguan community  \hspace{30pt}  & Nicaraguan   \hspace{10pt}  & $108$   \hspace{10pt}     & $1517$     \hspace{10pt}   & $14.05$ \hspace{10pt}     & $0.11$ & \cite{Jeremy2018}    \\    
UC Irvine messages               &uc-social & $1302$    & $19044$ & $29.3$     \hspace{10pt}   & $0.68$   &   \cite{KONECT}   \\ 
Online dating               &POK& $3562$    & $18098$ & $10.2$     \hspace{10pt}   & $0.78$   &   \cite{Makse}   \\  
\end{tabular}
\end{ruledtabular} 
\label{tabSI:data_desc} 
\end{table}


\end{widetext}

%


\bibliography{bibliography}
\end{document}